# Interfacial Synergy in Ag-Doped CuO-AgCl-g-C$_3$N$_4$ Composites for Efficient Charge Separation and Low-power Methylene Blue Degradation

Suresh Chandra Baral[1,2], Uttama Kumar Saint[3], Dilip Sasmal[1], Sradhanjali Lenka[1], Ashish Kalkal[4], A. Mekki[5,6], Sudhagar Pitchaimuthu[2*], and Somaditya Sen[1*]

[1]*Department of Physics, Indian Institute of Technology Indore, Indore, 453552, India*

[2]*Research Center for Carbon Solutions (RCCS), Institute of Mechanical, Process and Energy Engineering, School of Engineering and Physical Sciences, Heriot-Watt University, Riccarton Campus, Edinburgh, EH14 4AS, Scotland, United Kingdom*

[3]*Department of Electrical and Electronic Engineering, Ariel University, Ariel 4070000, Israel*

[4]*Institute of Biological Chemistry, Biophysics & Bioengineering, School of Engineering & Physical Sciences, Heriot-Watt University, Riccarton Campus, Edinburgh, EH14 4AS, Scotland, United Kingdom*

[5]*Department of Physics, King Fahd University of Petroleum and Minerals, Dhahran 31261, Saudi Arabia*

[6]*Interdisciplinary Research Center for Advanced Materials, King Fahd University of Petroleum and Minerals, Dhahran 31261, Saudi Arabia*

* Corresponding author: sens@iiti.ac.in (SS), S.Pitchaimuthu@hw.ac.uk (SP)

**Abstract:**

An Ag-doped CuO-AgCl-g-C$_3$N$_4$ heterostructure has been designed to achieve rapid Methylene Blue (MB) degradation through a synergistic photo-Fenton mechanism driven by low-power UV illumination. The composite integrates narrow-bandgap CuO, plasmonic Ag/AgCl, and visible-responsive g-C$_3$N$_4$ into a dual Z-scheme configuration that promotes efficient interfacial charge transfer while preserving strong redox potentials. Diffuse reflectance UV-Vis spectra ascertained the band-gap positions of the composite corresponding to those of its constituents: 2.9 eV (g-C$_3$N$_4$) and 1.42 eV (Ag-doped CuO-AgCl), indicating enhanced absorption and efficient charge carrier generation. BET analysis confirmed the presence of mesoporosity and revealed an effective surface area, ensuring the availability of abundant adsorption and reaction sites. A commercial 11 W UV irradiation was used for the photocatalytic test. Almost complete degradation of MB occurred within 10 min, following pseudo-first-order kinetics with a high apparent rate constant of 0.45 min$^{-1}$. The remarkable activity arises from the synergistic interplay of Fenton-like redox cycling and efficient photoinduced charge carrier generation and separation. In addition, it has been demonstrated that intentionally incorporated AgCl plays an active role as a plasmonic-semiconducting interface, strengthening charge separation and catalyst stability under neutral conditions, rather than acting as a passive chloride byproduct. Overall, by linking defect engineering, heterojunction design, and photo-Fenton synergy, this study establishes a low-power, catalytic platform offering a viable pathway towards sustainable dye wastewater remediation.

**Keywords:** Fenton-like degradation, Photo-fenton degradation, Low-power photocatalyst, Z-scheme configuration, g-C$_3$N$_4$,

1. **Introduction**

Water scarcity has become a continuous global issue due to the rapid increase in the human population, industrialization, and related pollution. The continuous discharge of synthetic dye wastewater, containing stubborn compounds such as textile dyes, poses a significant threat to both human and aquatic ecosystems. Moreover, these dyes disrupt the natural cleansing processes of water bodies, which are essential for photosynthesis by limiting light penetration. Among many textile dyes, Methylene Blue (MB), a typical cationic thiazine-based compound, exhibits toxicity and potential carcinogenic effects [1]. It is essential to employ treatment methods that go beyond surface-level decolorization and aim for complete breakdown into harmless end products to effectively tackle such pollutants [2].

Several treatment methods have been utilized to treat this dye wastewater. Particularly, advanced oxidation stands out for its ability to produce highly reactive oxygen species that break down organic pollutants into carbon dioxide and water [3]. Traditionally, the homogeneous Fenton process (Fe$^{2+}$/Fe$^{3+}$ chemistry in the presence of hydrogen peroxide (H$_2$O$_2$)) is recognized for its effectiveness in its degradation. However, it requires highly acidic conditions (typically around pH 2–3). Additionally, the process generates substantial amounts of iron sludge, which results in inefficient regeneration of Fe$^{3+}$ back to Fe$^{2+}$, thereby limiting its practical applicability [4]. Researchers have turned to heterogeneous Fenton-like systems that utilize solid transition-metal oxides (TMOs) to overcome these limitations. These metal oxides provided improved operational stability at near-neutral pH levels [5], [6], [7], facilitating easier separation and reuse after treatment [8], [9], [10], [11], [12]. In these systems, H$_2$O$_2$ is activated through redox interactions occurring at the surface of transition metal centers ($M^{n+}/M^{n+1}$), resulting in the generation of reactive oxygen species. Initially, $M^{n+}$ reacts with $H_2O_2$ to produce $M^{n+1}$ with the addition of hydroxyl radicals and hydroxide ions. Then, $M^{n+1}$ reduces back to $M^{n+}$ with light-induced charge carriers, ensuring continuous catalytic activity without generating sludge [13].

Among other TMOs, Copper-based oxides have gained significant attention for heterogeneous Fenton-like catalysis due to their relatively small band gap (ranging from 1.2 to 1.8 eV), good structural stability, and suitable redox ($Cu^{2+}/Cu^+$) chemistry [14], [15]. Recent kinetic analyses have demonstrated that $Cu_2O$ exhibits superior activity in decomposing $H_2O_2$. The superior activity is attributed to surface-level electron transfer mechanisms rather than bulk-phase reactions [16]. $Cu^+$ ions interact with $H_2O_2$, forming hydroxyl radicals at a rate constant of approximately $10^4$ $M^{-1} \cdot s^{-1}$. On the other hand, $Cu^{2+}$ is regenerated through reduction by either $H_2O_2$ or electrons generated under light irradiation, maintaining a continuous $Cu^{2+}/Cu^+$ redox cycle. This mechanism promotes the core Fenton-like behavior observed in copper-based catalysts.

Our research investigations have systematically advanced the CuO platform through defect engineering, morphology control, and cationic substitution. All these modifications have enhanced both photocatalytic and Fenton-like performance of CuO. Our first study focused on $Al^{3+}$- doping in CuO powders. The study demonstrated that the controlled substitution of smaller $Al^{3+}$ ions (0.53 Å) for $Cu^{2+}$ (0.71 Å) ions induces lattice contraction and creates Cu-vacancy defects that enhance hole mobility and p-type conductivity [17]. The optimized sample achieved an almost 99% degradation of MB within 6 minutes under a 0.79 W m$^{-2}$ LED and 98% under natural sunlight. The catalyst also demonstrated good stability, with negligible Cu/Al leaching and excellent reusability. Mechanistically, Al doping tuned the $Cu^{2+}/Cu^+$ equilibrium and maximized the dual functionality of $H_2O_2$, acting both as a radical precursor and an electron scavenger. This confirms the effectiveness of defect-modulated CuO as a low-power photocatalyst [18].

Our second study investigated CuO nanosheets synthesized via the hydrothermal method, utilizing NaOH as a morphology-directing agent [19]. The findings revealed a facet-dependent enhanced adsorption and charge transfer, confirming a structure-correlated catalytic activity. The catalyst resulted in 99% degradation of MB dye within 35 minutes and almost complete degradation within 70 minutes using an 11 W UV lamp and natural sunlight, respectively. A detailed thermodynamic analysis revealed the process to be endothermic and diffusion-coupled, with an apparent activation energy of ~56 kJ mol$^{-1}$. The catalyst demonstrated good stability, with minimal Cu leaching (~0.6 ppm), thereby validating the intrinsic stability and scalability of CuO.

Our third investigation introduced Fe$^{3+}$ into the CuO lattice, forming a dual redox system (Fe$^{3+}$/Fe$^{2+}$ ↔ Cu$^{2+}$/Cu$^{+}$) that simultaneously enhanced photocatalytic degradation and the oxygen evolution reaction (OER) [20]. The Fe-doped CuO achieved nearly complete MB removal within 50 min and exhibited improved electrochemical reversibility, confirming that coupled cationic redox cycling accelerates interfacial electron exchange and strengthens both photo- and electro-catalytic pathways. Similarly, Fang et al. demonstrated that CuO nanosheets prepared via an alkaline H$_2$O$_2$ route possess surface peroxo (–O–O–) groups that mediate visible-light H$_2$O$_2$ decomposition, generating •OH and •O$_2^-$ radicals [21]. Furthermore, Sun et al. emphasized that Cu$^+$-rich interfaces are the true active sites for H$_2$O$_2$ activation, with electron buffering by adjacent metallic or defect states crucial for sustained catalysis [22]. Furthermore, Bhowmick et al. reported a self-cyclic Mn$^{3+}$ + Cu$^{2+}$ ↔ Mn$^{4+}$ + Cu$^{+}$ redox system in Mn$_2$CuO$_4$ that achieved spontaneous neutral-pH dye degradation, underscoring the power of multivalent coupling in maintaining rapid radical generation [23]. Collectively, these findings establish that stabilized Cu$^+$ centers, surface peroxo intermediates, and synergistic redox coupling are the hallmarks of next-generation heterogeneous Fenton-like catalysts. However, most of these systems employ UV-based degradation without fully utilizing visible light, resulting in a limited penetration depth in aqueous media, higher energy consumption, and low solar utilization efficiency, thereby restricting large-scale, sustainable applications.

To harness visible light more effectively, an external semiconductor sensitizer is essential. Graphitic carbon nitride (g-C$_3$N$_4$), with a 2.7 eV band gap and conduction-band potential of −1.12 V vs NHE, can fulfill this role efficiently [24]. Beyond visible-light absorption, g-C$_3$N$_4$ performs dual redox functions: (i) *in situ* H$_2$O$_2$ generation through a two-electron O$_2$ reduction involving a 1,4-endoperoxide intermediate [25], and (ii) photoactivation of H$_2$O$_2$ to produce •OH radicals [26]. These complementary reactions make g-C$_3$N$_4$ a natural partner for CuO-based Fenton systems, coupling photon-driven charge excitation with chemical radical activation.

Building on this trajectory, the present work reports the synthesis and mechanistic evaluation of a ternary Ag-doped CuO-AgCl-g-C$_3$N$_4$ composite, engineered to unify broadband solar harvesting, accelerated charge dynamics, and efficient Fenton-like activation. The composite integrates three synergistic functionalities: (i) g-C$_3$N$_4$: visible-light harvester and in situ H$_2$O$_2$ generator, (ii) Ag-doped CuO-AgCl heterojunction: redox-active Fenton core providing Cu$^{2+}$/Cu$^+$ cycling and surface-mediated H$_2$O$_2$ activation, and (iii) Photo-responsive Ag@AgCl domains: plasmonic sensitizers and interfacial electron mediators that facilitate charge migration between CuO and g-C$_3$N$_4$ [27], [28]. This one-pot precipitation-assembly route employed here ensures intimate contact among CuO nanosheets, AgCl nanodomains, and g-C$_3$N$_4$ sheets, forming oxygen-vacancy-rich junctions that promote interfacial charge transport, achieving superior MB degradation under low-power UV light irradiation. This study demonstrates that controlled AgCl incorporation, far from being a residual byproduct of chloride precursors, can serve as an active plasmonic-semiconducting component that reinforces charge separation and catalytic durability under neutral conditions, providing a comprehensive framework that links defect engineering, heterojunction design, and photo-Fenton synergy for sustainable wastewater remediation.

## 2. Experimental section

### 2.1. Materials and Synthesis

#### 2.1.1. Synthesis of Ag-doped CuO-AgCl-g-C$_3$N$_4$ composite

The synthesis of the Ag-doped CuO-AgCl-g-C$_3$N$_4$ composite involved three steps: the preparation of the Ag-doped CuO-AgCl component, the exfoliation of g-C$_3$N$_4$, and the final solution-based assembly.

Ag-doped CuO-AgCl (AC) was prepared via a coprecipitation route in alkaline medium, followed by low-temperature hydrothermal treatment. In a 250 mL beaker, 3.00 g CuCl$_2$·2H$_2$O (17.6 mmol) was dissolved in 80 mL deionized (DI) water under magnetic stirring (60 min). AgCl (0.50 g, 3.49 mmol) was added, and the suspension was stirred for 30 min. Separately, 1.60 g of NaOH (40.0 mmol) was dissolved in 20 mL of DI water and then added dropwise (~10 min) to the Cu/Ag chloride mixture under vigorous stirring to reach a pH of ≥ 12, inducing the precipitation of mixed hydroxides. The slurry was aged for 30 minutes at room temperature and then transferred to a 150 mL Teflon-lined stainless-steel autoclave, with a total slurry volume of approximately 100 mL (approximately 67% of the liner volume). The autoclave was sealed and heated at 120°C for 24 h. After natural cooling to room temperature, the resulting dark-brown CuO–AgCl composite was collected by centrifugation, washed several times with DI water and ethanol to neutral pH, and dried at 60°C for 12 h. A mild anneal at 200°C for 2 h in air (ramp 2°C min$^{-1}$) was applied to enhance interparticle contact without promoting AgCl decomposition or Ag$^0$ reduction.

The total hydroxide added (0.0400 mol NaOH) slightly exceeded the amount required for complete conversion of $Cu^{2+}$ into $Cu(OH)_2$, i.e., 0.0352 mol, and partial surface reaction with AgCl (0.00349 mol). Under these alkaline-hydrothermal conditions, $Cu(OH)_2$ readily converts to CuO, whereas AgCl remains largely stable due to its low solubility and limited oxidative conversion to $Ag_2O$ [29], [30]. The obtained powder thus corresponds to a CuO–AgCl composite containing a minor amount of $Ag^+$ doped into the CuO lattice, consistent with XRD and compositional analyses.

Bulk $g-C_3N_4$ was prepared by thermal polymerization of urea. Urea (placed in a covered alumina crucible) was heated in air at 550°C for 2 h at a 5°C $min^{-1}$ ramp, yielding a yellow layered $g-C_3N_4$ powder with nanosheet morphology. To increase surface area and active sites, the bulk powder was chemically exfoliated. Specifically, $g-C_3N_4$ was dispersed in 1 M $H_2SO_4$ (solid–liquid ratio of 1 mg $mL^{-1}$) and stirred for 24 h at room temperature to protonate the interlayer nitrogen sites. This was followed by ultrasonication for 60 min to delaminate the sheets. The exfoliated $g-C_3N_4$ (gCN) was collected by centrifugation, washed repeatedly with DI water to near-neutral pH ($\approx$ 6–7), and dried at 60°C for 12 h.

The Ag-doped $CuO-AgCl-g-C_3N_4$ composite was prepared by co-dispersing the pre-synthesized CuO-AgCl powder and exfoliated $g-C_3N_4$ nanosheets in 70:30 (V/V) ethanol: DI water. Typical compositions included gCN-33 (0.05 g $g-C_3N_4$ + 0.10 g Ag-CuO-AgCl) and gCN-50 (0.10 g $g-C_3N_4$ + 0.10 g Ag-CuO-AgCl). Each component was ultrasonicated separately ($g-C_3N_4$, 30 min; Ag-CuO-AgCl, 10 min) prior to mixing under vigorous stirring. The dispersion pH was adjusted to 5.2 with 0.1 M $HNO_3$, promoting solution-mediated hetero-assembly between positively protonated $g-C_3N_4$ and the oxide surfaces (CuO/AgCl). The mixture was stirred for 2 hours with intermittent bath sonication (3 × 10 minutes). The product was recovered by centrifugation/filtration, washed with ethanol and DI water to neutral pH, dried at 60°C for 12 h, and gently post-annealed at 200°C for 2 h (air, 2°C $min^{-1}$) to consolidate interfacial bonding. The mildly acidic assembly environment (pH $\approx$ 5.2) balances surface charge attraction without dissolving AgCl or releasing $Ag^+$. Nitrate ions were used because they are weakly coordinating and leave no interfering residues, unlike $Cl^-$ (which could shift Ag equilibrium) or $SO_4^{2-}$ (which adsorbs strongly) [31], [32]. This strategy preserves the CuO–AgCl phase integrity while maximizing contact with $g-C_3N_4$, resulting in a stable ternary heterostructure suitable for photocatalytic evaluation [33], [34], [35].

### 2.2. Structural, Vibrational, and Electronic Characterization

X-ray diffraction (XRD) analysis of all the samples was performed using a Bruker D2 Phaser X-ray diffractometer with Cu-K$_\alpha$ radiation ($\lambda$ = 1.54 Å) to determine the structural details. Fourier transform infrared (FT-IR) spectra were recorded using a Spectrum Two FT-IR spectrometer (PerkinElmer). The surface morphology of all samples was characterized using Field Emission Scanning Electron Microscopy (FESEM, JEOL, JSM-7610 F). The pore size distribution, specific surface area, and pore volume were estimated from an $N_2$ adsorption-desorption measurement [Brunauer-Emmett-Teller (BET) surface area analysis] using a Quantachrome Autosorb iQ2 BET Surface Area and Pore Volume Analyzer. The valence state of cations and the oxygen content were analyzed from the X-ray Photoelectron Spectroscopy (XPS) data obtained using a Thermo-Scientific Escalab 250 Xi instrument equipped with a monochromatic Al K$_\alpha$ X-ray source (h$\nu$ = 1486.6 eV) operating at a power of 150 W and under UHV conditions in the range of $\sim 10^{-9}$ mbar. All spectra were recorded in hybrid mode, using electrostatic and magnetic lenses and an aperture slot of 300 μm × 700 μm. The wide and high-resolution spectra were acquired at fixed analyzer pass energies of 80 and 20 eV, respectively. A flood gun was used to compensate for charging effects.

### 2.3. Photocatalytic Properties Characterization

The photocatalysis was performed on commercially procured MB ($C_{16}H_{18}ClN_3S$, 99.9% purity, SRL chemicals, India). Hydrogen peroxide solution (30% w/v) ($H_2O_2$, 99.9% purity, Rankem, India) was procured as an oxidizing agent.

The photocatalytic activity of all the samples was evaluated using a commercially available low-pressure mercury-vapor discharge UV lamp that emits UV radiation with a wavelength of approximately 253.7 nm (Philips TUV 11W PLS) (Spectrum is given in the Supplementary, Figure S1). The light intensity was only 100 lx ≈ 0.79 Wm$^{-2}$ (i.e., 1266 times smaller than AM 1.5G Sun). A 10-ppm strength stock solution of pollutant MB dye was prepared by mixing 10 mg of the MB dye in 1 L of DI water. Catalyst concentrations of 0.25 mg/mL were prepared by adding 10 mg of precatalysts to 40 mL of MB dye solution in separate beakers. The obtained mixed solutions were stirred continuously in the dark for 60 minutes to achieve a good adsorption-desorption equilibrium. The concentration of the remaining dye was estimated from the time-dependent absorption spectra at different intervals recorded using a Research India UV-visible spectrophotometer. After each step of the photocatalytic degradation of samples, a 3 mL solution was extracted from the photoreacted solution. The photoreacted solutions were centrifuged properly to extract the catalyst nps. The reduction in intensity of the absorption spectra over time was studied for each sample. The photocatalytic degradation efficiency of the organic dyes was calculated using the following formula:

$$Degradation\ (\%) = [(A_0 - A_t)/A_0] * 100 \qquad [Eq.(A.1)]$$

where $A_0$ is the absorbance of the dye solution before the photo-irradiation, and $A_t$ is the absorbance of the solution after photo-irradiation for a specific time t.

## 3. Results and discussion

### 3.1. Structural analysis

The crystallographic features and phase composition of the synthesized samples were investigated via powder XRD, as presented in Figure 1. The pristine g-$C_3N_4$ (gCN) exhibits a single broad reflection centered at ≈ 27.3°, indexed to the (002) plane of graphitic carbon nitride (JCPDS 87-1526), which originates from the interlayer stacking of conjugated heptazine units [36]. The absence of any other reflections confirms the formation of the pure polymeric phase with a layered graphitic structure.

In contrast, the AC, gCN-33, and gCN-50 composites display multiple additional diffraction peaks, indicating the incorporation of crystalline CuO and AgCl phases. The intense peaks located at 35.5°, 38.7°, 48.7°, 53.5°, 58.3°, 61.5°, and 66.2° correspond to the (–111), (111), (–202), (020), (202), (–113), and (–311) planes of monoclinic CuO (tenorite, JCPDS 48-1548), confirming the presence of well-crystallized CuO nanoparticles within the composite [37]. Meanwhile, the relatively weaker reflections marked with "#" at 27.8°, 32.2°, 46.2°, 54.8°, 57.4°, 67.4°, and 74.4° can be indexed to the (111), (200), (220), (311), (222), (400), and (420) planes of face-centered cubic AgCl (JCPDS 31-1238), verifying the coexistence of AgCl nanocrystals [38]. No additional peaks associated with metallic Ag or other impurity phases are observed, indicating that Ag remains primarily in the AgCl lattice or as finely dispersed interfacial $Ag^0$ clusters below the XRD detection limit.

The characteristic (002) reflection of g-$C_3N_4$ persists in all composites but appears slightly broader and shifts marginally toward lower angles (≈ 25–26°), which suggests increased interlayer spacing and partial turbostratic disorder induced by the anchoring of CuO and AgCl nanoparticles between the g-$C_3N_4$ layers. Such peak broadening and shift provide evidence of strong interfacial coupling and lattice strain at the heterojunction interfaces. The simultaneous detection of CuO and

AgCl phases alongside the modified g-$C_3N_4$ stacking pattern unambiguously confirms the successful formation of a ternary CuO-AgCl-g-$C_3N_4$ heterostructure.

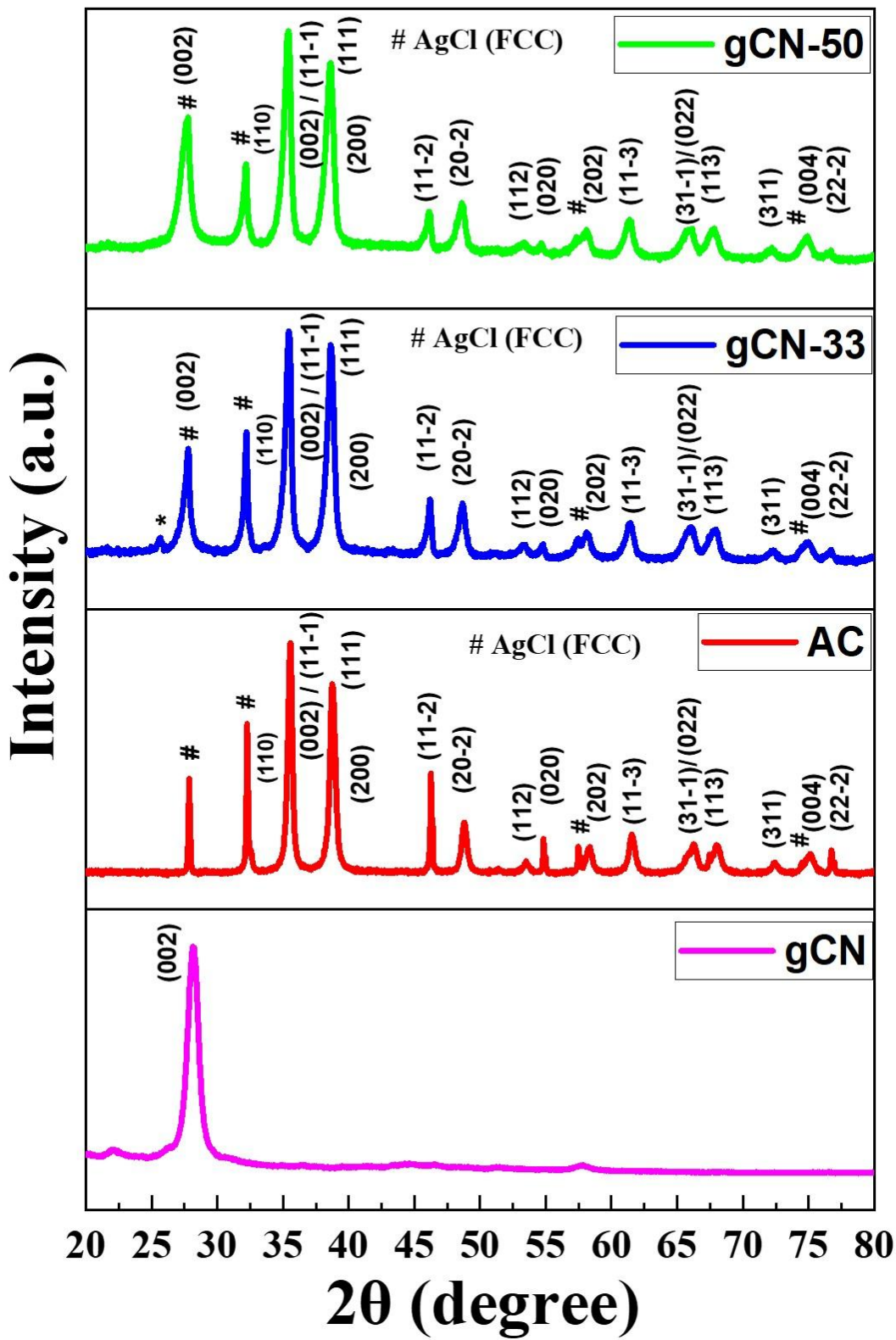

***Figure 1***. X-ray diffraction (XRD) patterns of pristine g-$C_3N_4$, CuO-AgCl (AC), and CuO-AgCl-g-$C_3N_4$ composites (gCN-33 and gCN-50). The pristine g-$C_3N_4$ shows a broad (002) reflection at ≈ 27.3°, characteristic of the interlayer stacking of conjugated heptazine units. The intense peaks observed at 35.5°, 38.7°, 48.7°, 53.5°, 58.3°, 61.5°, and 66.2° correspond to monoclinic CuO (JCPDS 48-1548), while the weaker reflections marked "#" at 27.8°, 32.2°, 46.2°, 54.8°, 57.4°, 67.4°, and 74.4° are assigned to FCC AgCl (JCPDS 31-1238). The coexistence of CuO and AgCl reflections together with the slightly broadened and down-shifted g-$C_3N_4$ (002) peak (~25–26°) confirms the successful formation of a ternary CuO-AgCl-g-$C_3N_4$ heterostructure, where interfacial coupling among the three phases induces lattice strain and expanded interlayer spacing.

Complementary FTIR analysis (Figure 2) further supports these observations. The spectra exhibit the typical g-$C_3N_4$ vibrations around 810 cm$^{-1}$ (triazine ring breathing) and 1200-1650 cm$^{-1}$ (C–NH–C and C–N(–C)=C stretching), confirming retention of the heptazine framework [39]. A new broad absorption band in the 450-600 cm$^{-1}$ region corresponds to Cu-O stretching, consistent with XRD results [40]. The broad 3000-3400 cm$^{-1}$ band associated with N–H and O–H stretching shows increased intensity in the composites, revealing improved surface hydroxylation and amine terminations [41], [42], [43].

Overall, the XRD and FTIR results collectively demonstrate that CuO and AgCl nanoparticles are successfully anchored onto the g-$C_3N_4$ matrix, forming a structurally integrated heterojunction system. The intimate contact between these components is expected to facilitate interfacial charge transfer, contributing to the enhanced photocatalytic and photoelectrochemical performance discussed in the subsequent sections.

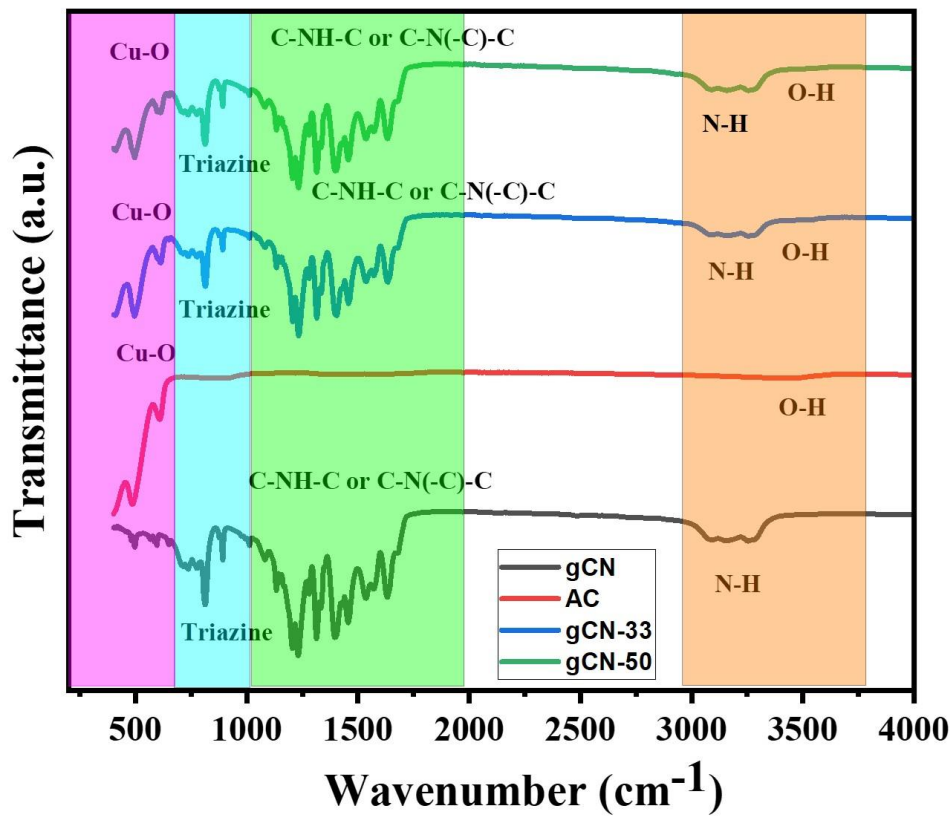

**Figur*e 2. FTIR spectra of pure g-C$_3$N$_4$, CuO–AgCl (AC), and CuO–AgCl–g-C$_3$N$_4$ composites (gCN-33 and gCN-50). The broad absorption band around 450–600 cm$^{-1}$ corresponds to the Cu–O lattice vibration of CuO, confirming successful formation of the oxide. The sharp band near 810 cm$^{-1}$ arises from the triazine ring-breathing mode, while multiple peaks in the 1200–1650 cm$^{-1}$ range are attributed to C–NH–C and C–N(–C)=C stretching of the heptazine framework. The broad feature between 3000–3400 cm$^{-1}$ originates from N–H and O–H stretching vibrations, indicating surface –NH$_x$/–OH functionalities. The coexistence of these characteristic bands verifies the preservation of the g-C$_3$N$_4$ network and the formation of an interfacial bonded CuO–AgCl–g-C$_3$N$_4$ heterostructure.*

### 3.2. Morphological characterization

The surface morphology and microstructural features of the synthesized samples were examined by FESEM, as shown in Figure 3. The pristine g-C$_3$N$_4$ exhibits a lamellar and wrinkled nanosheet morphology, which is characteristic of the exfoliated graphitic structure derived from thermal polycondensation of urea [44], [45]. The thin, loosely stacked sheets offer a high surface area and abundant edge sites that serve as anchoring points for the subsequent nucleation of metal and metal oxide nanoparticles.

In contrast, the CuO–AgCl (AC) sample exhibits a dense assembly of CuO nanoplates and short nanorods, intertwined with smaller AgCl nanoparticles, resulting in a hierarchically rough architecture. This morphology arises from the controlled coprecipitation synthesis route, in which $Ag^+$ and $Cu^{2+}$ precursors undergo concurrent hydrolysis and chloride-mediated transformation. The close spatial distribution of AgCl on CuO surfaces ensures strong electronic coupling between the two components.

Upon integration with g-$C_3N_4$, the composites (gCN-33 and gCN-50) exhibit distinct morphological evolution. In gCN-33, the CuO–AgCl nanostructures are uniformly dispersed and intimately anchored onto the g-$C_3N_4$ nanosheets, creating a porous, flower-like network with well-defined interfaces. When the g-$C_3N_4$ content is increased to 50 wt% (gCN-50), the structure becomes more continuous and sheet-dominated, with partial coverage of CuO–AgCl domains within the extended g-$C_3N_4$ framework. This morphological consolidation indicates enhanced interfacial coupling between the semiconductor phases, consistent with the XRD-observed lattice modulation and FTIR-identified Cu–O–C/N bonding features.

Overall, the morphological analyses corroborate that CuO and AgCl nanoparticles are successfully anchored onto the g-$C_3N_4$ sheets, forming a structurally stable and intimately bonded ternary heterostructure. The abundant interfacial junctions and hierarchical surface architecture are expected to enhance charge separation and provide more accessible active sites for subsequent catalytic processes.

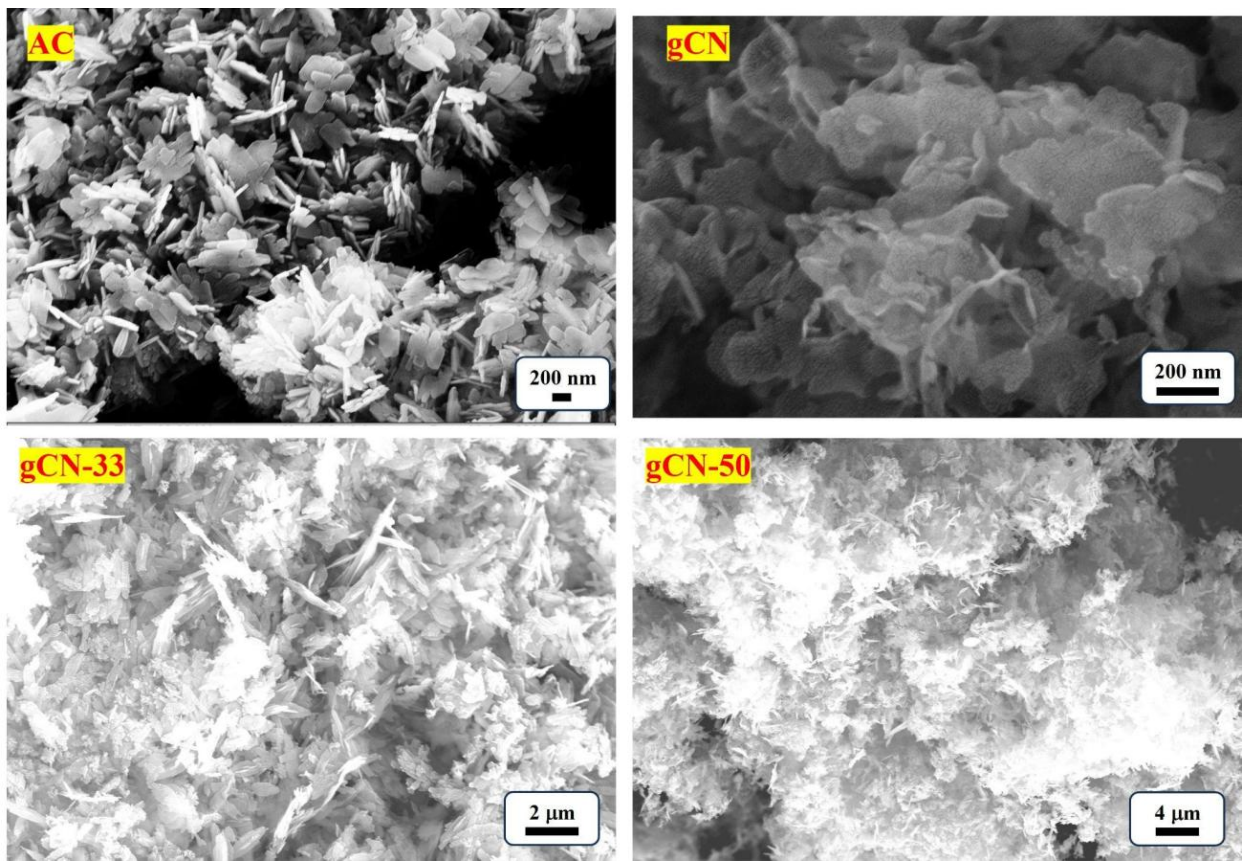

*Figure 3. FESEM images of (a) CuO-AgCl (AC), (b) pristine g-$C_3N_4$, (c) gCN-33, and (d) gCN-50 composites at different magnifications. The pristine g-$C_3N_4$ shows a stacked, wrinkled nanosheet morphology characteristic of the exfoliated graphitic structure. The AC sample exhibits densely packed CuO nanoplates/nanorods, with AgCl particles uniformly distributed on their surfaces, resulting in a hierarchical microstructure. After coupling with g-$C_3N_4$, the CuO–AgCl nanostructures are homogeneously anchored onto the carbon nitride sheets, leading to increased roughness and interfacial contact in gCN-33. Increasing the g-$C_3N_4$ content to 50 wt% (gCN-50) yields a more sheet-dominated and interconnected architecture, where CuO–AgCl features remain embedded within an expanded g-$C_3N_4$ network. The progressive transition from lamellar g-$C_3N_4$ to a three-phase CuO–AgCl–g-$C_3N_4$ heterostructure highlights enhanced interfacial contact beneficial for charge transfer and catalytic activity.*

### 3.3. Optical Analysis

Diffuse-reflectance UV-Vis spectra were converted to absorbance using the Kubelka-Munk function and subsequently plotted in Tauc form, $[\alpha h\nu]^2$ vs $h\nu$, assuming direct-allowed transitions [46]. As shown in Figure 4, pristine g-$C_3N_4$ shows a sharp absorption edge in the near-UV, with an optical gap of ~2.9 eV, characteristic of the π-π* transition in the tri-s-triazine framework [47]. After chemical exfoliation, the nanosheets retain this near-UV onset but display slightly enhanced visible-tail absorption due to disorder-induced states.

Ag-doped CuO-AgCl (AC) exhibits broad absorption across the UV-vis-NIR region, with a strong visible component arising from the narrow-gap CuO domain and Ag/AgCl-related defect levels. The Tauc intercept is ~1.4 eV, consistent with a defect-rich p-type CuO matrix [48].

The g-$C_3N_4$-AC composites (gCN-33 and gCN-50) exhibit a combination of these behaviours. Both spectra retain the g-$C_3N_4$ edge in the near-UV while showing markedly increased visible absorption derived from the AC component. The apparent Tauc onsets decrease in the order gCN-50 (~2.75 eV) > gCN-33 (~2.6 eV) > AC (~1.42 eV). This systematic red shift indicates stronger interfacial coupling and more efficient sub-gap transitions as the AC fraction increases [49]. Importantly, composites typically do not form a single, new band gap. Instead, they exhibit multiple optical transitions originating from the individual components (g-$C_3N_4$ and CuO/AgCl) together with interfacial or defect-assisted absorptions. The observed reduction in the apparent optical gap for gCN-33 and gCN-50reflects both (i) the contribution of the AC domain and (ii) new interfacial states created at the CuO/AgCl-g-$C_3N_4$ junction, rather than a true bandgap narrowing.

Between the two composites, gCN-33 shows slightly stronger visible absorption and a smaller apparent gap than gCN-50. This suggests that a higher CuO/AgCl fraction (at a fixed Ag doping level) enhances electronic communication at the interface and increases the effective joint density of states near the band edge [50], [51]. When considered together with the BET trend (gCN > gCN-50 > gCN-33 > AC), the optical data indicates that gCN-33 provides the best balance between visible-light harvesting and accessible surface area, which is a favourable combination for rapid photo-Fenton degradation where both photon capture and interfacial charge transfer govern the reaction kinetics.

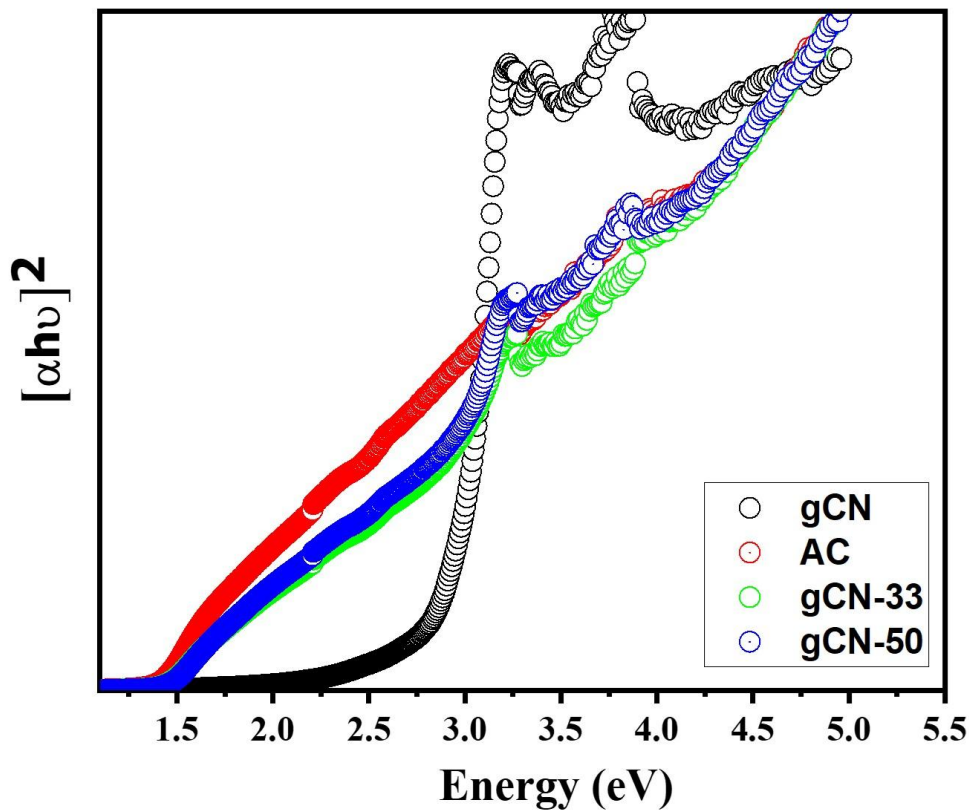

*Figure 4. Tauc plots [αhν]² vs photon energy (eV) for g-C₃N₄ (gCN), Ag-doped CuO–AgCl (AC), and their composites (gCN-33, gCN-50). Band-gap energies were estimated from linear fits to the near-edge region. The composites show pronounced redshifts and enhanced visible absorption relative to gCN, with gCN-33 displaying the lowest apparent gap among the gCN-loaded samples, indicative of stronger interfacial coupling and defect-assisted transitions.*

### 3.4. Compositional Analysis

The XPS survey spectrum of the Ag-doped CuO-AgCl sample was collected with Al Kα radiation. The survey confirms the presence of Cu and O as the dominant surface species, with an adventitious C 1s signal. Ag 3d and Cl 2p contributions are weak in the survey spectrum but are resolved in the corresponding high-resolution regions. No other elements apart from the ones mentioned were detected. High-resolution XPS spectra were taken for the Ag-doped CuO-AgCl sample to assess its surface composition and chemical states. Peak fitting analysis utilizing the

XPSPEAK41 software with a Shirley background was carried out to resolve the chemical states and different species of each element. The Cu 2p spectrum exhibits doublet features ($2p_{3/2}$ and $2p_{1/2}$) arising from the strong spin-orbit coupling with a spin-orbit splitting energy ($\Delta E_{SOS}$) of ~19.8 eV. The $2p_{3/2}$ and $2p_{1/2}$ XPS peaks exhibit a relative area ratio of 2:1, which arises from the spin-orbit coupling-induced degeneracy [52]. The $2p_{3/2}$ components appear around 933.70 eV (main) and 936.42 eV (secondary), and the corresponding $2p_{1/2}$ partners are derived at 953.50 eV and 956.22 eV, respectively. Apart from that, the spectrum consists of multiple pronounced shake-up satellite peaks around ~941–944 eV, arising from ligand to metal charge-transfer configuration of $Cu^{2+}$ species, and the $2p_{1/2}$ counterparts of those satellites appear around ~960–962 eV [53], [54]. The weak shoulder around ~936.4 eV can be attributed to surface Cu–OH species ($Cu(OH)_x/CuO_x(OH)_y$) arising from partial hydroxylation of CuO, in good agreement with the dominant –OH feature in the O 1s spectrum (530.43 eV, 46.0%). The Ag 3d doublet around 367.03/373.39 eV ($\Delta E_{SOS} \approx 6.36$ eV) corresponds to $AgCl/Ag^+$ species, while the O 1s region deconvolutes into lattice $O^{2-}$ (529.44 eV, 31.8%), surface –OH (530.43 eV, 46.0%), and adsorbed $H_2O/CO_x$ (531.99 eV, 22.2%). The high fraction of hydroxyl oxygen and the Cu–OH feature collectively indicate a hydroxyl-rich, defect-activated surface, favorable for interfacial redox reactions and catalytic activity [55], [56], [57].

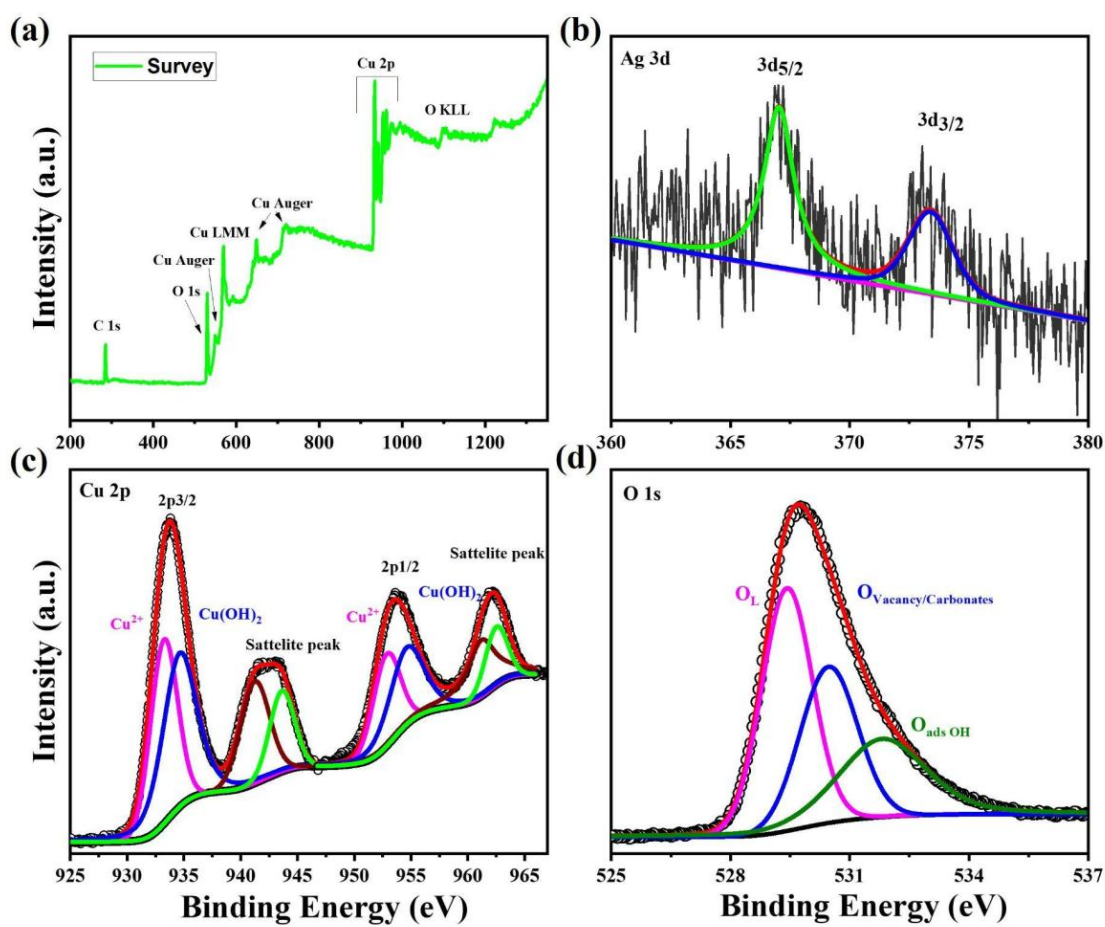

*Figure 5. XPS spectra of the Ag-doped CuO–AgCl composite. (a) Survey spectrum confirming the presence of Cu, O, Ag, and C elements, along with distinct Auger features (Cu LMM and O KLL). (b) Ag 3d region showing the $3d_{5/2}$ and $3d_{3/2}$ doublet at 367.0 eV and 373.4 eV, respectively, characteristic of $Ag^+$ species in the AgCl environment. (c) Cu 2p spectrum exhibiting main $Cu^{2+}$ peaks at 933.7 eV ($2p_{3/2}$) and 953.5 eV ($2p_{1/2}$), with satellite peaks near 941-944 eV and 960-962 eV that confirm the $Cu^{2+}$ oxidation state. A weak shoulder at ~936.4 eV indicates surface Cu-OH species formed due to partial hydroxylation of CuO. (d) O 1s spectrum deconvoluted into lattice oxygen ($O_L$, 529.4 eV), surface hydroxyl/defect oxygen ($O_{V/-OH}$, 530.4 eV), and adsorbed oxygen or carbonate species ($O_{ads}$, 532.0 eV), highlighting a hydroxyl-rich, defect-active surface favorable for interfacial redox reactions.*

### 3.5. Surface Area Analysis

The $N_2$ adsorption–desorption isotherms of the prepared samples (Figure 6) show type IV isotherms with $H_3$-type hysteresis loops, confirming the presence of mesoporous surface morphology in all the samples. All the isotherms exhibit a steep adsorption at higher relative pressures, due to multilayer adsorption. The adsorbed volume of $N_2$ for pure $g-C_3N_4$ (Figure 6a) shows very large values compared to other composites, consistent with inter-layer and inter-particle mesoporosity. The specific surface area ($S_{BET}$) for pure exfoliated $g-C_3N_4$ was found to be 59.55 $m^2/g$, attributed to the layered morphology that provides abundant accessible mesopores. In contrast, the Ag-doped CuO–AgCl (AC, Figure 6b) sample exhibits a comparatively lower $S_{BET}$ value of 12.88 $m^2/g$, which is consistent with the denser, aggregated oxide–halide particles that limit gas adsorption. However, the $H_3$-type hysteresis suggests the persistence of inter-aggregate mesoporosity. In gCN-33 heterocomposite (Figure 6c), which consists of 33 wt% $g-C_3N_4$ and 67 wt% AC, the specific surface area increases to 39.86 $m^2/g$. The introduction of the layered $g-C_3N_4$ fraction prevents tight aggregation of CuO–AgCl particles and introduces additional interfacial voids. When the $g-C_3N_4$ content is raised to 50 wt% (gCN-50, Figure 6d), with an equal weight ratio of gCN and AC, the surface area slightly increases further to 40.95 $m^2/g$, indicating a more balanced dispersion of AC particles within the layered $g-C_3N_4$ structure. However, the marginal difference in surface area between gCN-33 and gCN-50 suggests that gCN-33 provides the most effective combination of porosity and active-site accessibility for catalytic reactions.

The distinct textural characteristics of the synthesized materials are expected to strongly influence both dye adsorption and subsequent photo-Fenton degradation behavior [58], [59]. The high surface area and mesoporous nature of $g-C_3N_4$ provide abundant active sites and diffusion channels, favoring enhanced dye uptake and interfacial contact with reactive oxygen species. Upon incorporating Ag-doped CuO–AgCl, the surface area decreases slightly; however, the introduction of heterojunction interfaces and redox-active Cu/Ag sites is anticipated to accelerate $H_2O_2$ activation and charge transfer, compensating for the loss in porosity. Among the composites, gCN-33, with its balanced surface area and hierarchical pore structure, is expected to achieve an optimal compromise between adsorption capacity and catalytic activity. In contrast, the binary Ag-doped CuO–AgCl sample despite its larger pore diameter, likely exhibits limited dye adsorption and slower kinetics due to reduced surface area. Overall, the progressive modulation of surface texture and interfacial architecture is predicted to play a decisive role in governing adsorption equilibrium, electron–hole separation, and reactive radical generation during the photo-Fenton process.

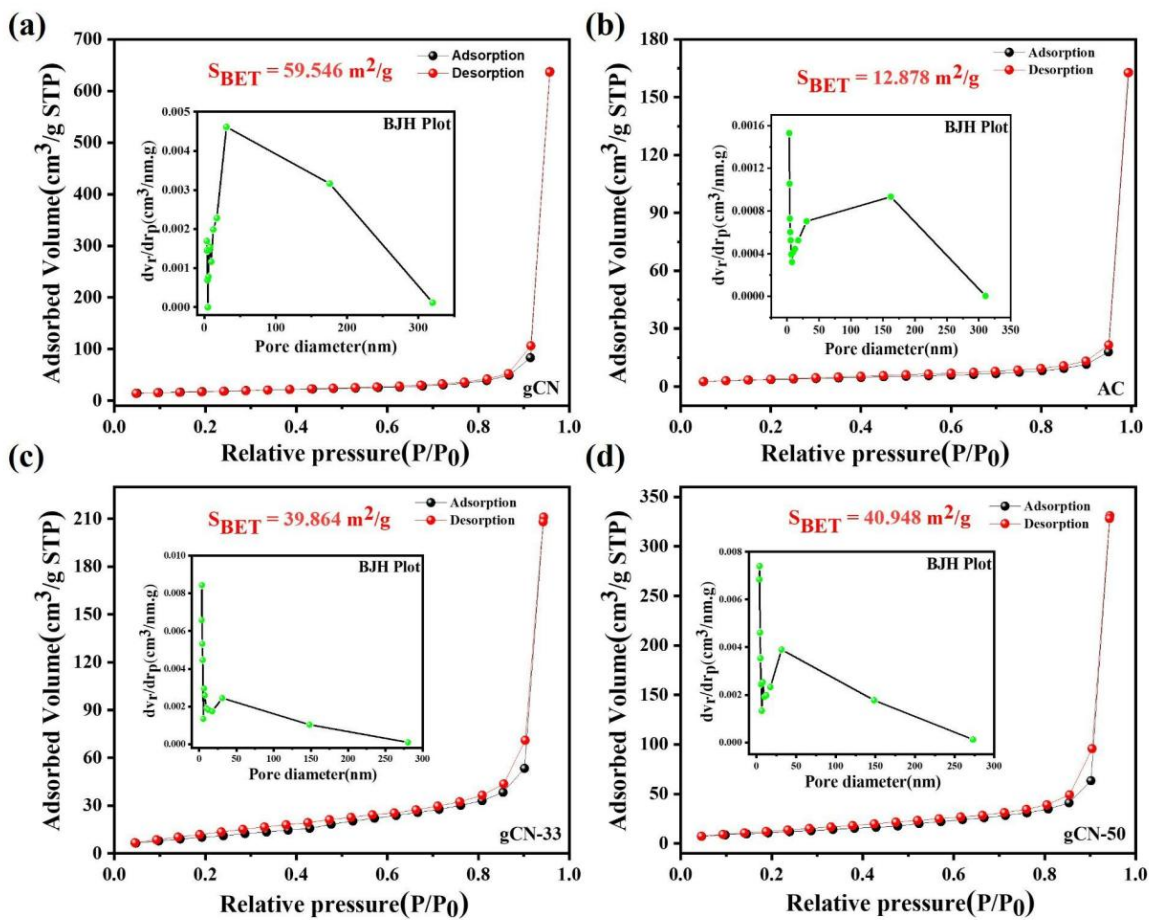

*Figure 6. $N_2$ adsorption-desorption isotherms and corresponding BJH desorption pore-size distributions of (a) g-$C_3N_4$, (b) Ag-doped CuO–AgCl (AC), (c) gCN-33, and (d) gCN-50. All samples exhibit type IV isotherms with $H_3$-type hysteresis loops, confirming the formation of mesoporous frameworks through inter-particle voids and stacked nanosheets. The BET surface areas are 59.55, 12.88, 39.86, and 40.95 $m^2/g$ for gCN, AC, gCN-33, and gCN-50, respectively. The insets show BJH desorption pore-size distributions, revealing dominant mesopore diameters of 20–60 nm. The pronounced increase in adsorbed volume at $P/P_0 \approx 0.99$ is attributed to capillary condensation within interparticle mesovoids. Among the composites, gCN-33 may exhibit the highest photo-Fenton catalytic activity, attributed to its balanced surface area, hierarchical porosity, and efficient interfacial charge-transfer pathways.*

### 3.6. Catalytic and Photocatalytic measurement

Conventionally, photocatalytic activity is evaluated using high-intensity UV or xenon lamps (250-300 W) or under direct sunlight irradiation [5], [60], [61], [62]. However, such light sources are expensive and energy-intensive. In the present study, an effort was made to assess the photocatalytic performance of the synthesized Ag-doped CuO-AgCl-g-$C_3N_4$ composite for the degradation of MB dye, a representative cationic thiazine dye widely used in the textile and printing industries. The photocatalytic experiments were conducted using a low-cost, commercially available low-pressure mercury-vapor discharge UV lamp that emits shortwave UV radiation at 253.7 nm (Philips TUV 11W PLS). To start with, the experiment was conducted in steps.

### 3.7. Dark Adsorption Study (Adsorption-Desorption Equilibrium)

To accurately evaluate the intrinsic photocatalytic activity, an adsorption-desorption equilibrium study was performed in the dark using a 10 ppm MB solution under continuous stirring. Prior to light illumination, the catalyst-dye suspensions were magnetically stirred in the dark for 60 min to attain equilibrium in the adsorption-desorption process between MB molecules and the catalyst surface. The adsorption efficiency was calculated using Eq.(A.1).

The pure g-$C_3N_4$ sample exhibited the highest adsorption capacity with 28.62 % MB removal after 60 min, attributed to its high surface area and abundant nitrogen-containing surface sites that promote electrostatic attraction toward the cationic $MB^+$ molecules. In contrast, Ag-doped CuO-AgCl exhibited a relatively lower adsorption efficiency of 8.77**%**, possibly due to its hydrophobic surface and limited affinity toward MB at near-neutral pH.

The Ag-doped CuO-AgCl/g-$C_3N_4$ composites exhibited intermediate adsorption behavior, with gCN-33 and gCN-55 removing 17.2% and 17.8% of MB, respectively, during the dark adsorption process. The moderate adsorption capacity of these composites can be ascribed to partial coverage of the g-$C_3N_4$ surface by CuO-AgCl nanoparticles, which reduces the number of accessible –NH/–OH adsorption sites but introduces additional interfacial heterogeneity and surface roughness. Such balanced adsorption is favorable for photocatalysis, as it ensures sufficient contact between the dye and catalyst without blocking photon absorption sites [63].

### 3.8. Dark Fenton-like Catalytic Degradation

After attaining adsorption–desorption equilibrium, 0.18 mL of 30 % (W/V) $H_2O_2$ (corresponding to 40 mM final concentration) was introduced into the 40 mL MB solution to evaluate the dark Fenton-like catalytic activity. The systems were kept under continuous stirring in the absence of light for an additional 60 min, and the temporal change in MB concentration was monitored spectrophotometrically at 664 nm.

In the dark, the decomposition of H₂O₂ on the catalyst surface can generate reactive oxygen species (ROS), primarily hydroxyl radicals(•OH), through surface-mediated redox reactions involving transition metal centers [Eq. (B.1) to Eq. (B.2)]:

$$M^{n+} + H_2O_2 \rightarrow M^{(n+1)+} + \bullet OH + OH^-  \quad [Eq.~(B.1)]$$

$$M^{(n+1)+} + H_2O_2 \rightarrow M^{n+} + \bullet OOH + H^+  \quad [Eq.~(B.2)]$$

Where M represents $Cu^{2+}/Cu^+$ or $Ag^+/Ag^0$ surface species acting as redox-active sites, such reactions occur even in the absence of light, although typically at a slower rate than photo-assisted processes.

The total MB removal efficiencies after 60 min were 46.6 % for g-C₃N₄, 12.1 % for Ag-doped CuO-AgCl, 99.0 % for gCN-33, and 96.0 % for gCN-50. After correcting for their respective dark adsorption contributions (28.6 %, 8.8 %, 17.2 %, and 17.8 %), the actual catalytic degradation achieved through H₂O₂ activation is 18.0 % for g-C₃N₄, 3.3 % for Ag-doped CuO-AgCl, 81.8 % for gCN-33, and 78.2 % for gCN-50, respectively.

The markedly enhanced activity of gCN-33 and gCN-50 under dark conditions indicates an efficient Fenton-like catalytic mechanism, wherein the redox-active Cu and Ag species promote the decomposition of H₂O₂ into reactive oxygen species (•OH and •O₂⁻) through surface-mediated electron transfer cycles ($Cu^{2+}/Cu^+$ and $Ag^+/Ag^0$). The intimate interfacial contact between CuO-AgCl nanoparticles and g-C₃N₄ nanosheets accelerates the regeneration of low-valent metal sites, sustaining continuous ROS formation even in the absence of light.

In contrast, the low degradation efficiencies observed for pristine g-C₃N₄ and binary Ag-doped CuO-AgCl suggest limited ability to activate H₂O₂ without photoexcitation. Overall, the near-complete MB removal achieved by gCN-33 (approximately 99%) within 60 minutes highlights its exceptional dark Fenton-like catalytic performance, demonstrating that optimized heterojunction interfaces can effectively drive redox reactions independently of external illumination.

### 3.9. Photocatalytic Degradation of MB under Low-power UV Light

Following the adsorption-desorption equilibrium study and Dark Fenton-like Catalytic study, photocatalytic degradation of MB solution (10 ppm in DI water) was performed under a low-pressure mercury-vapor discharge UV lamp that emits shortwave UV radiation at 253.7 nm

(Philips TUV 11W PLS), in the presence of 10 mM $H_2O_2$. Prior to illumination, all suspensions were maintained in the dark for 60 min to evaluate both **a**dsorptive removal and dark Fenton-like degradation.

During this dark stage, total MB removal reached 46.2 % for g-$C_3N_4$, 13.5 % for Ag-doped CuO-AgCl, 79.2 % for gCN-33, and 70.9 % for gCN-50. These values include contributions from surface adsorption and catalytic decomposition of $H_2O_2$ via metal-mediated Fenton-like reactions. Upon switching on the light, rapid photoactivation was observed. Within only 10 min (sampled every 2 min), total MB removal increased to 81.5 % for g-$C_3N_4$, 16.8 % for Ag-doped CuO-AgCl, 99 % for gCN-33, and 97.5 % for gCN-50. The abrupt enhancement in activity confirms the photo-Fenton synergy, in which light excitation of g-$C_3N_4$ generates photocarriers while CuO and AgCl/Ag domains facilitate interfacial charge transfer and continuous generation of reactive oxygen species (OH and $•O_2^-$) via $H_2O_2$ activation.

The near-complete decolorization achieved by gCN-33 within 10 min under such extremely low-intensity light highlights the outstanding efficiency and energy-saving potential of the Ag-doped CuO-AgCl-g-$C_3N_4$ heterostructure.

The MB degradation kinetics followed the rate law of a pseudo-first-order reaction, expressed as:

$$\ln(C_0/C_t) = k_{aPP} t \quad [\text{Eq. (C.1)}]$$

Where $C_0$ and $C_t$ represent the MB concentrations at 0 min and t min (after dark adsorption correction), respectively, and $k_{app}$ is the apparent rate constant. Linear correlations ($R^2 > 0.98$) were obtained for all samples, confirming first-order reaction kinetics as followed by the photo-Fenton process [Figure 7b]. The calculated rate constants show that the optimized gCN-33 exhibited the highest activity with $k_{app} = 0.446$ min$^{-1}$, nearly 4 times faster than pure g-$C_3N_4$ (0.108 min$^{-1}$) and > 121 times faster than Ag-doped CuO-AgCl (0.00368 min$^{-1}$). The slightly lower $k_{app}$ for gCN-50 (0.246 min$^{-1}$) compared to gCN-33 suggests that moderate CuO-AgCl loading yields optimal interfacial charge transfer and ROS generation efficiency.

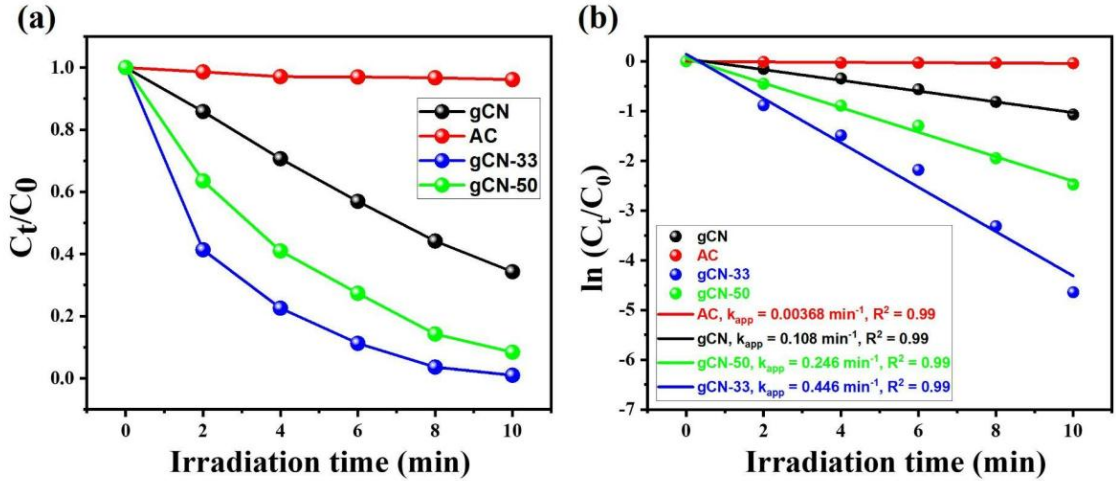

*Figure 7. Photocatalytic degradation of Methylene Blue (MB, 10 ppm) over different catalysts in the presence of 10 mM $H_2O_2$ under low-power UV light. (a) Temporal variation of normalized concentration ($C_t/C_0$) with irradiation time. (b) Corresponding pseudo-first-order kinetic plots [$ln(C_0/C_t)$ vs time]. The apparent rate constants ($k_{app}$) follow the order gCN-33 (0.446 min$^{-1}$) > gCN-50 (0.246 min$^{-1}$) > g-$C_3N_4$ (0.108 min$^{-1}$) > AC (0.00368 min$^{-1}$), indicating the superior photo-Fenton activity of the Ag-doped CuO-AgCl-g-$C_3N_4$ composites under low-power UV light irradiation.*

### 3.10. Proposed Synergistic Photo-Fenton Mechanism under Light Illumination

The enhanced photo-Fenton performance of the Ag-doped CuO-AgCl-g-$C_3N_4$ composites under low-power (11 W) UV irradiation (λ = 254 nm) arises from the cooperative interplay among photoexcitation, surface redox cycles, and interfacial charge transfer within the multi-component heterostructure (Figure 8). Despite the low incident energy, the well-aligned band structure and redox-active sites ensure efficient generation of reactive oxygen species (ROS) through combined photocatalytic and Fenton-like pathways.

#### 3.10.1. Photoexcitation and charge-carrier migration

Upon 254 nm illumination, AgCl ($E_g \approx 3.2$ eV) and g-$C_3N_4$ ($E_g \approx 2.9$ eV) absorb photons to generate electron-hole pairs:

$$AgCl + h\nu \rightarrow e^- (AgCl\_CB) + h^+ (AgCl\_VB) \quad [Eq. (D.1)]$$

$$g\text{-}C_3N_4 + h\nu \rightarrow e^- (g\text{-}C_3N_4\_CB) + h^+ (g\text{-}C_3N_4\_VB) \quad [Eq. (D.2)]$$

Meanwhile, CuO ($E_g \approx 1.42$ eV) can also undergo excitation or act as an efficient electron acceptor due to its narrow band gap. Metallic $Ag^0$ nanoparticles, generated in situ from partial photoreduction of AgCl, function as electron mediators bridging the conduction bands of CuO, AgCl, and g-$C_3N_4$ [64], [65]. The close interfacial contact enables a dual Z-scheme configuration, where the photogenerated electrons (CuO/AgCl) recombine with holes (g-$C_3N_4$) at $Ag^0$, while retaining high-energy electrons in g-$C_3N_4$ and holes in CuO/AgCl. This mechanism preserves the strong redox potentials required for ROS formation and effectively suppresses bulk recombination.

### 3.10.2. Radical generation via photocatalytic, photolytic, and Fenton-like routes

The separated charge carriers initiate a series of oxidative and reductive reactions:

I. Oxidation by AgCl/CuO valence-band holes

$$h^+ + OH^-/H_2O \rightarrow {}^\bullet OH \qquad [Eq.\ (D.3)]$$

II. Reduction of $O_2$ by g-$C_3N_4$ conduction-band electrons

$$e^- + O_2 \rightarrow O_2^- \rightarrow HO_2{}^\bullet \rightarrow H_2O_2 \qquad [Eq.\ (D.4)]$$

III. Cu/Ag-assisted $H_2O_2$ activation (Fenton-like cycle)

$$Ag^0 + H_2O_2 \rightarrow Ag^+ + {}^\bullet OH + HO^- \qquad [Eq.\ (D.5)]$$

$$Cu^{2+}/Cu^+ + H_2O_2 \rightarrow {}^\bullet OH + Cu^+/Cu^{2+} \qquad [Eq.\ (D.6)]$$

Step 1, $\quad [Cu(II)] + H_2O_2 \rightleftharpoons [Cu(II)]\ldots.H_2O_2 \rightarrow [Cu(I)] + {}^\bullet OOH + H^+ \qquad [Eq.\ (D.7)]$
$\qquad\qquad [Cu(I)] + H_2O_2 \rightleftharpoons [Cu(I)]\ldots.H_2O_2 \rightarrow [Cu(II)] + {}^\bullet OH + HO^- \qquad [Eq.\ (D.8)]$

Step 2, $\qquad\qquad {}^\bullet OOH + H_2O_2 \rightarrow H_2O + {}^\bullet OH + O_2 \qquad [Eq.\ (D.9)]$
$\qquad\qquad {}^\bullet OH + H_2O_2 \rightarrow H_2O + {}^\bullet OOH \qquad [Eq.\ (D.10)]$

IV. Direct UV photolysis of $H_2O_2$

$$H_2O_2 + h\nu \rightarrow 2\ {}^\bullet OH \qquad [Eq.\ (D.11)]$$

Thus, even at 11 W lamp power, multiple ROS pathways operate concurrently, photo-induced charge reactions [Eq. (D.1) to Eq. (D.4)], catalytic $H_2O_2$ decomposition via $Cu^+/Cu^{2+}$ and $Ag^0/Ag^+$ cycles [Eq. (D.5) to Eq. (D.10)], and direct photolysis of $H_2O_2$ [Eq. (D.11)]-leading to sustained radical generation under mild conditions.

### 3.10.3. Synergistic photo-Fenton oxidation of MB

The highly oxidative •OH and $HO_2•/•O_2^-$ radicals produced in the above steps attack adsorbed Methylene Blue (MB) molecules, breaking the thiazine ring and converting intermediates into $CO_2$, $H_2O$, and inorganic ions. The synergy arises from: (i) Efficient charge separation in the dual Z-scheme architecture (CuO ↔ Ag ↔ g-$C_3N_4$), (ii) Simultaneous $H_2O_2$ activation by Cu and Ag surface redox couples, and (iii) Supplementary UV photolysis enhances ROS yield. Together, these processes enable nearly complete MB mineralization under low-intensity UV light, proving that optimized heterostructural and redox coupling can achieve high activity without high-power illumination [66], [67].

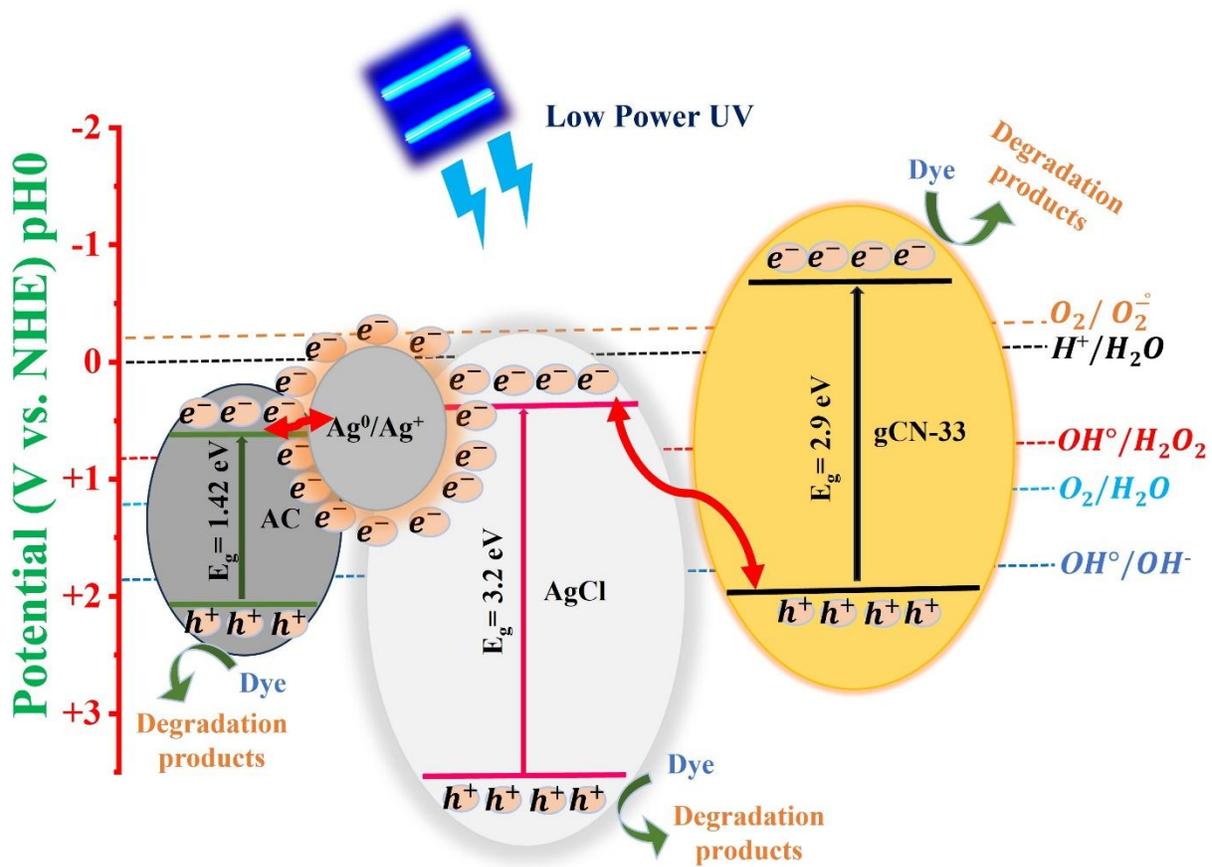

*Figure 8: Schematic illustration of the proposed synergistic photo-Fenton mechanism over the Ag-doped CuO-AgCl–g-C$_3$N$_4$ composite under 11 W, 254 nm UV irradiation. The diagram illustrates photoexcitation of AgCl, CuO, and g-C$_3$N$_4$. Photo-generated electrons migrate from the AgCl conduction band to the gCN-33 valence band for recombination, making the first Z-scheme pathways. At the same time, photogenerated electrons migrate through Ag$^0$ bridges, creating pathways for Fenton-like H$_2$O$_2$ activation via Cu$^+$/Cu$^{2+}$ and Ag$^0$/Ag$^+$ cycles, producing •OH radicals for rapid dye degradation. While energetic electrons in g-C$_3$N$_4$ create •O$_2^-$ radicals, holes from AgCl and AC are responsible for direct oxidation, collectively contributing to overall degradation.*

## 4. Conclusion and Future Perspectives

A series of Ag-doped CuO-AgCl-g-C$_3$N$_4$ composites were successfully synthesized and evaluated for the degradation of MB through a combination of adsorption, dark Fenton-like, and photo-Fenton processes. The catalysts were tested under a low-pressure mercury-vapor discharge UV lamp that emits shortwave UV radiation at 253.7 nm (Philips TUV 11W PLS), demonstrating an energy-efficient strategy for dye degradation and wastewater remediation.

The equilibrium adsorption studies revealed moderate MB uptake (17–18 %) for the gCN composites, confirming the presence of accessible active sites and favorable dye-surface interactions. Upon the addition of 10 mM H$_2$O$_2$, pronounced Fenton-like activity was observed, achieving 79.2% and 70.9% MB removal for gCN-33 and gCN-50, respectively, attributed to the synergistic redox participation of Cu$^{2+}$/Cu$^+$ and Ag$^+$/Ag$^0$ couples. When illuminated under UV light, the degradation proceeded rapidly, yielding 99.9% (gCN-33) and 97.5% (gCN-50) MB removal within 10 minutes. Kinetic analysis revealed pseudo-first-order behavior, with apparent rate constants following the order gCN-33 (0.446 min$^{-1}$) > gCN-50 (0.246 min$^{-1}$) > g-C$_3$N$_4$ (0.108 min$^{-1}$) > AC (0.00368 min$^{-1}$).

Mechanistic investigations established that enhanced performance arises from a dual Z-scheme (g-C$_3$N$_4$ ↔ CuO ↔ AgCl/Ag) charge-transfer pathway. In this configuration, photoexcited electrons in g-C$_3$N$_4$ and holes in CuO/AgCl are spatially separated yet energetically preserved, while metallic Ag$^0$ functions as an electron mediator, facilitating interfacial charge recombination of low-energy carriers and sustaining continuous H$_2$O$_2$ activation into •OH and •O$_2^-$ radicals. The cooperative effect between photoinduced charge separation and Fenton-like redox cycling enables rapid mineralization of MB even under very weak illumination.

Overall, this study demonstrates that Ag-doped CuO-AgCl-g-$C_3N_4$ is an exceptionally active and energy-efficient photo-Fenton catalyst, capable of achieving nearly complete dye degradation under ambient or indoor lighting conditions. The findings highlight its potential for low-power, sustainable wastewater treatment and provide a platform for designing multifunctional heterostructured catalysts for environmental remediation.

Future investigations will focus on applying this low-power photo-Fenton system to real textile wastewater containing mixed contaminants and evaluating catalyst reusability and structural stability over successive cycles. Incorporation of the composite into continuous-flow or membrane-photoreactor configurations may further enable scalable operation using natural or indoor light. In addition, operando spectroscopic analyses (e.g., in situ XPS, Raman, or XAS) will be pursued to elucidate the dynamic Cu/Ag redox transformations that govern the synergistic photo-Fenton mechanism. These efforts are expected to advance the development of next-generation, energy-efficient photocatalytic systems for sustainable environmental purification.

**Acknowledgment:**

The authors would like to acknowledge the Department of Science and Technology (DST), Government of India, for providing the funds (DST/TDT/AMT/2017/200). SCB would like to thank DST INSPIRE for providing fellowships (IF190617). SCB would also like to thank the Commonwealth Scholarship Commission (CSC) UK, which provides a split-site scholarship offering a research visit and support at Heriot-Watt University. The authors would like to acknowledge the Sophisticated Instrument Centre (SIC) facilities at IIT Indore for providing the BET surface area analyzer. The authors also acknowledge the DST, Government of India, New Delhi, India, for providing a FIST instrumentation fund to the discipline of Physics at IIT Indore to purchase a Raman Spectrometer (Grant Number SR/FST/PSI-225/2016).

**Declaration of generative AI and AI-assisted technologies in the manuscript preparation process:**

During the preparation of this work, the authors utilized generative AI to improve language and readability. After using this tool, the authors reviewed and edited the content as needed and took full responsibility for the content of the published article.